\documentclass{jfm}

\usepackage{color}

\usepackage{graphicx}
\usepackage{natbib}
\usepackage{amsmath}

\ifCUPmtlplainloaded \else
  \checkfont{eurm10}
  \iffontfound
    \IfFileExists{upmath.sty}
      {\typeout{^^JFound AMS Euler Roman fonts on the system,
                   using the 'upmath' package.^^J}%
       \usepackage{upmath}}
      {\typeout{^^JFound AMS Euler Roman fonts on the system, but you
                   dont seem to have the}%
       \typeout{'upmath' package installed. JFM.cls can take advantage
                 of these fonts,^^Jif you use 'upmath' package.^^J}%
      }
  \else
  \fi
\fi

\ifCUPmtlplainloaded \else
  \checkfont{msam10}
  \iffontfound
    \IfFileExists{amssymb.sty}
      {\typeout{^^JFound AMS Symbol fonts on the system, using the
                'amssymb' package.^^J}%
       \usepackage{amssymb}%
       \let\le=\leqslant  \let\leq=\leqslant
       \let\ge=\geqslant  \let\geq=\geqslant
      }{}
  \fi
\fi

\ifCUPmtlplainloaded \else
  \IfFileExists{amsbsy.sty}
    {\typeout{^^JFound the 'amsbsy' package on the system, using it.^^J}%
     \usepackage{amsbsy}}
    {\providecommand\boldsymbol[1]{\mbox{\boldmath $##1$}}}
\fi

\title{Swinging and tumbling of elastic capsules in shear flow}

\author[S. Kessler, R. Finken and U. Seifert]%
{S.\ns K\ls E\ls S\ls S\ls L\ls E\ls R,%
  \ns R.\ns F\ls I\ls N\ls K\ls E\ls N,%
 \ns \and U.\ns S\ls E\ls I\ls F\ls E\ls R\ls T}

\affiliation{II. Institut f{\"u}r Theoretische Physik, Universit{\"a}t
  Stuttgart, 70550 Stuttgart, Germany}

\date{27 June 2007}

\newcommand{\vol}{V}
\newcommand{\lenscale}{R_0}

\newcommand{\poisson}{\nu}
\newcommand{\dimbend}{\tilde{\bend}}
\newcommand{\dimspont}{\tilde{C}_0}
\newcommand{\dimshear}{\chi}

\newcommand{\membrane}{\mathcal{M}}
\newcommand{\refmembrane}{\mathcal{M_\text{ref}}}

\newcommand{\infindex}{\infty}
\newcommand{\inindex}{\text{i}}
\newcommand{\outindex}{\text{o}}
\newcommand{\bendindex}{\bend}
\newcommand{\elastindex}{\text{el}}
\newcommand{\induced}{\text{ind}}

\newcommand{\visc}{\eta}
\newcommand{\viscout}{\visc^\outindex}
\newcommand{\viscin}{\visc^\inindex}
\newcommand{\visccont}{\epsilon}

\renewcommand{\vec}[1]{\boldsymbol{#1}}
\newcommand{\ten}[1]{\boldsymbol{#1}}

\renewcommand{\div}[1]{\vec{\nabla}\cdot{#1}}
\newcommand{\grad}[1]{\vec{\nabla}{#1}}
\newcommand{\laplace}[1]{\Delta{#1}}

\newcommand{\ort}{x}
\newcommand{\shape}{r}
\newcommand{\refshape}{R}

\newcommand{\lt}{\vartheta}
\newcommand{\lp}{\varphi}

\newcommand{\sr}{r}
\newcommand{\sth}{\theta}
\newcommand{\sph}{\phi}

\newcommand{\surgrad}{\nabla^\text{S}{}}

\newcommand{\basis}{e}
\newcommand{\normal}{n}
\newcommand{\metric}{g}
\newcommand{\fl}{a}

\newcommand{\refbasis}{E}
\newcommand{\refnormal}{N}
\newcommand{\refmetric}{G}
\newcommand{\reffl}{A}

\newcommand{\kronecker}{\delta}

\newcommand{\norm}[1]{\left|{#1}\right|}

\newcommand{\curv}{k}
\newcommand{\mean}{H}
\newcommand{\gauss}{K}

\newcommand{\tr}[1]{\text{tr}{\,{#1}}}

\newcommand{\lagr}{\varepsilon}

\newcommand{\strain}{\lagr}

\newcommand{\surdil}{J}

\newcommand{\exten}{\lambda}

\newcommand{\energy}{\mathcal{H}}
\newcommand{\refenergyden}{h}
\newcommand{\force}{f}

\newcommand{\bend}{\kappa}
\newcommand{\spont}{C_0}
\newcommand{\lamemu}{\mu}
\newcommand{\lamelambda}{\lambda}

\newcommand{\shearrate}{\dot{\gamma}}
\newcommand{\shear}{\shearrate}

\newcommand{\vel}{u}
\newcommand{\p}{p}
\newcommand{\stress}{T}

\newcommand{\y}[2]{Y^{#2}_{#1}}
\newcommand{\cy}[2]{Y^{\ast{#2}}_{#1}}

\newcommand{\vy}[2]{\vec{Y}^{#2}_{#1}}
\newcommand{\vphi}[2]{\vec{\Phi}^{#2}_{#1}}
\newcommand{\vpsi}[2]{\vec{\Psi}^{#2}_{#1}}

\newcommand{\lamb}{U}
\newcommand{\lambp}[2]{\vec{\lamb}^{p}_{{#1},{#2}}}
\newcommand{\lambphi}[2]{\vec{\lamb}^{\phi}_{{#1},{#2}}}
\newcommand{\lambchi}[2]{\vec{\lamb}^{\chi}_{{#1},{#2}}}

\newcommand{\bw}{b}

\newcommand{\cart}{e}
\newcommand{\spherical}{e}

\newcommand{\mat}{A}

\newcommand{\tenpseudo}{\ten{A}^{-1}}

\newcommand{\deform}{D}

\newcommand{\longax}{L}
\newcommand{\shortax}{S}

\newcommand{\incl}{\beta}
\newcommand{\tank}{\alpha}
\newcommand{\phase}{\delta}

\begin{document}

\maketitle
\begin{abstract} \label{sec:abstract} 
The deformation of an elastic micro-capsule in an infinite shear flow is
studied numerically using a spectral method. The shape of the capsule and
the hydrodynamic flow field are expanded into smooth basis functions.
Analytic expressions for the derivative of the basis functions permit the
evaluation of elastic and hydrodynamic stresses and bending forces at
specified grid points in the membrane. Compared to methods employing a
triangulation scheme, this method has the advantage that the resulting
capsule shapes are automatically smooth, and few modes are needed to
describe the deformation accurately.  Computations are performed for
capsules both with spherical and ellipsoidal unstressed reference shape.
Results for small deformations of initially spherical capsules coincide with
analytic predictions. For initially ellipsoidal capsules, recent
approximative theories predict stable oscillations of the tank-treading
inclination angle, and a transition to tumbling at low shear rate. Both
phenomena have also been observed experimentally. Using our numerical
approach we could reproduce both the oscillations and the transition to
tumbling. The full phase diagram for varying shear rate and viscosity ratio
is explored. While the numerically obtained phase diagram qualitatively
agrees with the theory, intermittent behaviour could not be observed within
our simulation time. Our results suggest that initial tumbling motion is
only transient in this region of the phase diagram.
\end{abstract}

\section{Introduction}
The dynamics of soft objects such as drops, capsules and cells in flow
represents a long-standing problem in science and engineering, but has
received increasing interest recently, in particular due to its relevance to
biological, medicinal and microfluidic applications. This problem is
challenging from a theoretical point of view, because the shape of these
objects is not given \emph{a priori}, but determined dynamically from a
balance of interfacial forces with fluid stresses. Improved experimental
methods have revealed intriguing new dynamical shape transitions due to the
presence of shear flow. The phenomenology of the dynamical behaviour depends
distinctively on the specific soft object immersed into the flow with fluid
bilayer vesicles and elastic microcapsules as prominent classes.

Fluid bilayer vesicles assume a stationary tank-treading shape in linear shear
flow, if there is no viscosity contrast between interior and exterior fluid
\citep{kraus1996}. If the interior fluid or the membrane becomes more viscous,
a transition to a tumbling state can occur
\citep{biben2003a,beaucourt2004b,rioual2004a,noguchi2004,noguchi2005,vlahovska2007}.
Tank-treading was observed experimentally in infinite shear flow
\citep{haas1997,kantsler2005a} and for vesicles interacting with a rigid wall
\citep{lorz2000,abka02}, where a dynamic lift occurs
\citep{seifert1999b,cantat1999a,sukumaran2001,beaucourt2004a}. The
tank-treading to tumbling transition was observed for the first time
convincingly in a recent experiment \citep{kantsler2006a}. In addition to the
tank-treading to tumbling transition, a vacillating or breathing motion was
predicted theoretically \citep{misbah2006} and observed experimentally
\citep{kantsler2006a} and in simulations \citep{noguchi2007}. The theoretical
description has been expanded recently beyond first order in the shear rate
\citep{lebedev2007}.

In contrast to fluid vesicles, microcapsules exhibit a finite shear
elasticity, since their membrane is chemically or physically cross-linked.
This includes both artificial polymerised capsules \citep{walter2001} and red
blood cells (RBCs), whose membrane is composed of an incompressible lipid
bilayer underlined by a thin elastic cytoskeleton \citep{mohandas1994}.  The
resistance to shear leads to qualitatively different behaviour, such as
preventing the prolate to oblate shape transition in viscous fluid vesicles in
channel flow \citep{noguchi2005}. Perhaps most surprisingly, it also leads to
qualitatively different instabilities like wrinkling first observed
experimentally on polymerised capsules \citep{walter2001}, which has to be
distinguished from the transient creasing formation observed later on fluid
vesicles \citep{kantsler2007}. The formation of the short length scale
wrinkles is driven by compressive stress imposed on the membrane by the flow,
while the selection of the short length scale is due to a balance between
elastic stresses and bending forces \citep{finken2006}.

When the unstressed initial shape of the cell is not spherical, material
elements of the membrane are deformed when displaced from their initial
position. This shape memory, suggested for RBCs by \citet{fischer2004}, leads
to a oscillation of the inclination angle superimposed on the tank-treading
motion and an intermittent regime between tank-treading and tumbling
\citep{skotheim2007,abkarian2007}. For a review of the tank-treading behaviour
of soft capsules in shear flow, we refer the reader to the first two chapters
of \citet{pozrikidis2003book}.

Analytic descriptions of the rather complex motion of capsules and vesicles is
only possible for asymptotic cases, e.g. in the quasispherical limit
\citep{barthes-biesel1980,barthes-biesel1981,seifert1999a,misbah2006,finken2006,lebedev2007,vlahovska2007},
or by restricting the number of degrees of freedom
\citep{keller1982,rioual2004a,skotheim2007}. One therefore has to resort to
numerical methods for more complex geometries.

For the dynamics of vesicles existing solvers which treat the flow at a
continuum level either employ a discrete triangulation scheme
\citep{kraus1996} or phase field models \citep{biben2003a}. An alternative
route was taken by \citet{noguchi2004,noguchi2005}, where the membrane is
dynamically triangulated and the flow is modelled by discrete effective fluid
particles.

Numerical simulations of capsules were first performed in an axisymmetric
geometry \citep{li1988,leyrat-maurin1993,leyrat-maurin1994}.
\citet{pozrikidis1995} developed a method for simulating three-dimensional
deformations of initially spherical capsules in shear flow using a boundary
element formulation. This method was later refined by \citet{ramanujan1998},
who also observed oscillations of the inclination angle for ellipsoidal
capsules. However, their method was plagued by numerical instabilities for
high and low deformations due to the degradation of the grid. Further
improvement of the boundary element method allows the stable simulation of
tank-treading and tumbling motion of highly flattened capsules only by
numerically smoothing the surface \citep{pozrikidis2003a}. Part of the
numerical difficulties might be due to the evaluations of bending moments:
Calculating the local mean curvature requires taking second derivatives of the
shape functions, which become inaccurate using finite difference schemes.
Newer approaches, such as the spectral boundary algorithms proposed for
droplets \citep{Wang2006} and solid particles \citep{pozrikidis2006a},
therefore use higher order basis functions on the triangulated surface. While
these methods are very versatile, the details of these methods are rather
complex. Suitable interfacial smoothing is still needed to ensure numerical
stability of the method \citep{Wang2006}.

It is therefore the purpose of this paper to augment these approaches with a
global spectral method. In this method the shape of the capsule is expanded
into a set of smooth basis functions \citep{boyd2001}. This has the advantage
that the resulting shape is automatically globally smooth, which reduces the
discretisation error especially in higher derivatives, such as the local mean
curvature. Since the derivatives of the basis functions are analytically
known, it is easy to evaluate the elastic tensions and bending moments at a
grid of collocation points. These marker points are material points of the
underlying connected membrane. Rather than treating the hydrodynamics in a
boundary layer formulation, we expand the Stokes flow similarly in terms of
smooth basis functions. Requiring force balance at the collocation points
yields the equation of motion of the membrane. This scheme is used to
systematically explore the dynamic behaviour of capsules in shear flow,
focusing on initially non-spherical capsules as considered in the analysis by
\citet{skotheim2007}.  Although the overall phase behaviour of the capsules is
captured qualitatively by their model, we could not observe the predicted
intermittent behaviour. An analysis of the capsule dynamics suggests that the
initial tumbling motion is only a transient towards a stable tank-treading
motion.

This paper is organized as follows: In section \ref{sec:problem}, after
introducing notions of differential geometry and elasticity, we define the
problem rigorously. In section \ref{sec:spectral}, we develop the spectral
algorithm to calculate the dynamics of an elastic capsule. After extensive
testing for analytically known limit cases in section \ref{sec:spherical}, we
apply our method to ellipsoidal capsule in section \ref{sec:ellipsoidal}. Our
findings are summarised in section \ref{sec:discussion}. In the appendix, we
recall the relevant differential geometry for deformed capsules.

\section{Problem formulation} \label{sec:problem}
\begin{figure}
  \begin{center}
   \includegraphics[width=0.6\linewidth]{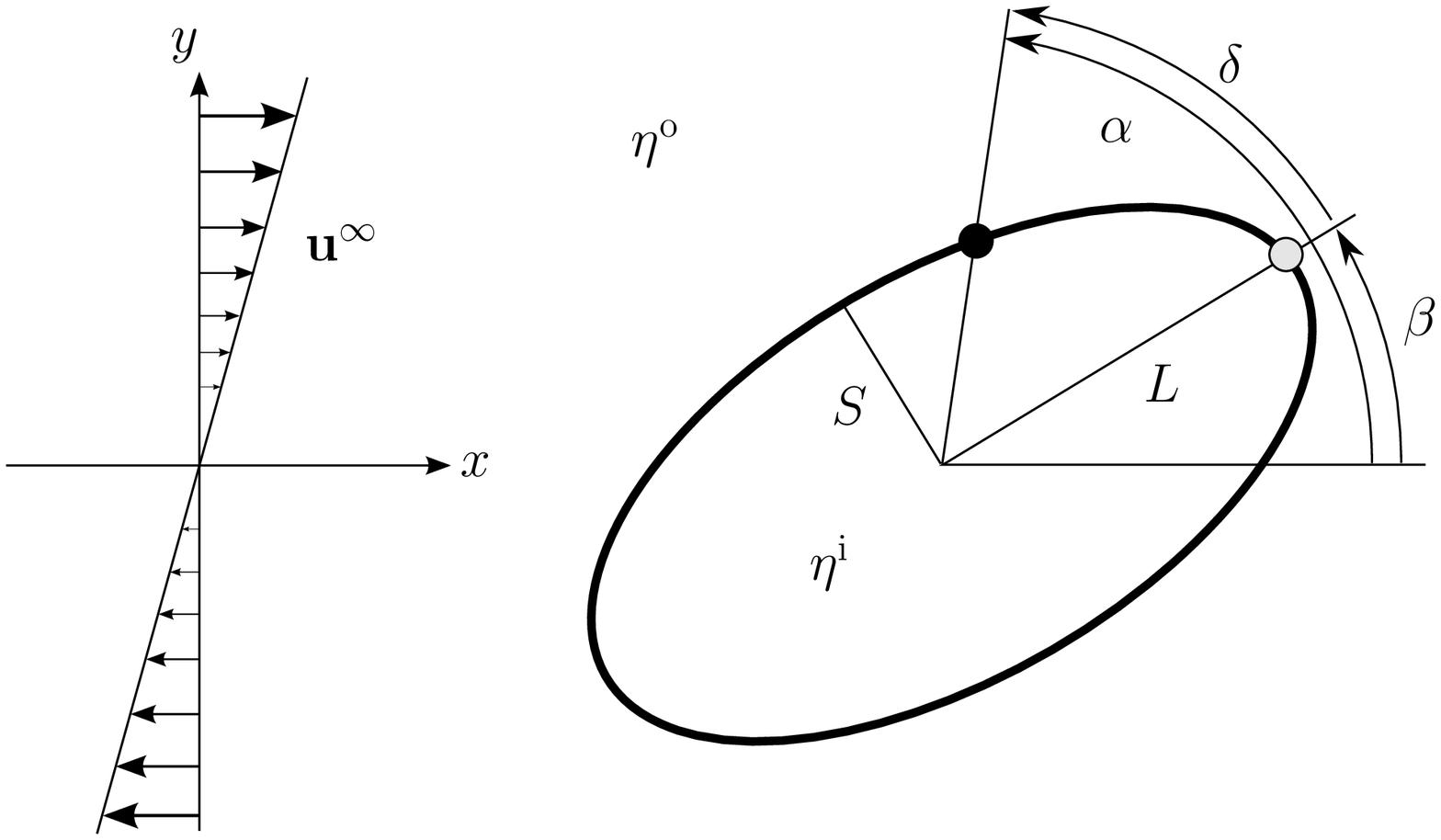}
   \caption{Elastic capsule in hydrodynamic flow. --- The viscosity of the
     outer flow and the inner fluid are $\viscout$ and $\viscin$ respectively.
     Long and short axes of the deformed capsule are denoted by $\longax$ and
     $\shortax$. The inclination angle $\incl$ measures the angle between the
     direction of maximal elongation and that of the shear flow
     $\vec{\vel}^\infty=\shear y \vec{\cart}_x$. The angle defined by a tracer
     particle on the membrane compared to the flow direction is denoted by
     $\tank$.}
   \label{fig:capsule2}
  \end{center}
\end{figure}

We consider the dynamics of a closed capsule that is embedded in an infinite
ambient flow with viscosity $\viscout$ (see figure \ref{fig:capsule2}). The
elastic membrane encloses a second fluid with a different viscosity $\viscin$,
defining the dimensionless viscosity contrast
\begin{equation}
 \visccont \equiv \viscin/\viscout \,. 
 \label{eq:visc_cont}
\end{equation}
In the absence of the capsule we assume a prescribed external flow
$\vec{\vel}^\infindex(\vec{\ort})$. Because of its small thickness we
consider the membrane as a two dimensional interface that separates the two
fluids. On the typical length scales considered inertial effects of the
membrane are negligible.

\subsection*{Strain and Curvature}
In order to describe the two dimensional membrane, which is embedded in three
dimensional space, we recall some important terms of differential geometry
\citep{frankel1997, marsden1983}. For mathematical details of the quantities
used here, we refer the reader to the appendix. A comprehensive summary of
interfacial properties in the context of membranes in hydrodynamic flow can be
found in \citet{Pozrikidis2001}.

Since we consider closed membranes with the topology of a sphere $S^2$, we can
label the material points of some reference membrane by spherical coordinates
$(\lt,\lp)$. Note that for an arbitrarily deformed membrane the material point
labelled by the Lagrange coordinates $(\lt,\lp)$ will be moved, and
$(\lt,\lp)$ will not remain spherical coordinates.

The location of the membrane $\membrane$ at time $t$ is given by the shape
function $\vec\shape(\lt,\lp;t)$. Length and angles on the membrane are
measured by the first fundamental form or metric tensor $\ten{\metric}$ with
covariant components $\metric_{ij}$ (\ref{eq:metric}). The second fundamental
form or extrinsic curvature tensor $\ten{\curv}$ with covariant components
$\curv_{ij}$ (\ref{eq:curvature}) measures how the unit normal vector of the
surface changes its direction, when one moves along the membrane. The mean
curvature $\mean$ is defined as the arithmetic mean of the principle
curvatures, which are both the eigenvalues of the curvature tensor
$\ten{\curv}$ and the inverse of the principle curvature radii
(\ref{eq:mean}). In our convention (\ref{eq:curvature}), the mean curvature
$\mean=2/r$ of a sphere with radius $r$ is positive. First and second
fundamental forms completely fix the shape of the given membrane and therefore
contain all information about the membrane shape.

In order to describe a deformation and an elastic response, we have to specify
an unstressed reference membrane $\refmembrane$ given by the shape function
$\vec\refshape(\lt,\lp)$. The corresponding metric is denoted by
$\ten{\refmetric}$ and defined analogously to the metric tensor
$\ten{\metric}$. The Lagrangian strain tensor $\ten{\lagr}$ is in covariant
components $\lagr_{ij}$ given by half the difference of metric $\metric_{ij}$
and reference metric $\refmetric_{ij}$ (\ref{eq:lagr}) and measures the change
in length elements of the membrane upon deformation
\cite[appendix,][]{marsden1983}. The strain tensor $\ten{\strain}$ holds all
information about the deformation and will be used to define an elastic energy
density.

Finally, the surface dilation $\surdil$ (\ref{eq:surdil}) measures how an
infinitesimal patch of area $d\reffl$ on the reference membrane is changed
upon deformation into the patch of area $d\fl$ (\ref{eq:area}) and can be
expressed by the ratio of determinants of metric and reference metric
(\ref{eq:det_metric}).

\subsection*{Constitutive laws, energy, force, stress}
The deformation of the membrane from its unstressed shape costs energy, which
can be quantified by the underlying constitutive law. In general we consider
resistance against shear, dilation, and bending. Several elastic models for
thin shells and membranes are considered in the literature. A short overview
is found in \citet{barthes-biesel2002} and in the first two chapters of
\citet{pozrikidis2003book}. These references directly connect the deformation
with stresses and bending moments. We prefer to derive the constitutive law
from the elastic energy, as outlined in \cite{marsden1983}.

Deformation from the reference shape $\vec{\refshape}(\lt,\lp)$ to a shape
$\vec{\shape}(\lt,\lp;t)$ costs energy $\energy[\vec{\shape}]$. We only
consider constitutive laws derived from an energy density
$\refenergyden[\vec{\shape}]$, i.e.
\begin{equation}
  \energy[\vec{\shape}] = \int\limits_\refmembrane d\reffl \, \refenergyden[\vec{\shape}]\,.
\end{equation}
Variation of the shape $\vec{\shape}$ by $\delta \vec{\shape}$ while leaving
the reference shape fixed induces a variation of the total free energy
\begin{equation}
  \delta \energy = \int_\refmembrane d\reffl\, \frac{\delta \energy[\vec{\shape}]
    }{\delta \vec{\shape}} \cdot \delta \vec{\shape} \equiv - \int_\membrane
  d\fl\, \vec{\force} \cdot \delta \vec{\shape} \,,
  \label{eq:energy_variation}
\end{equation}
which defines the elastic surface force density $\vec{\force}$ on the membrane
by a functional derivative
\begin{equation}
  \vec{\force} \equiv - \frac{1}{\surdil} \frac{\delta \energy[\vec{\shape}]}{\delta \vec{\shape}} \,.
\end{equation}

We now specialise to deformation energies, which can be written as the sum of
a purely elastic and a bending part,
\begin{equation}
  \energy[\vec{\shape}] = \energy_{\elastindex}[\vec{\shape}] + \energy_{\bendindex}[\vec{\shape}] \,.
\end{equation}

To illustrate the method, we use the specific constitutive law for the elastic
free energy
\begin{equation}
  \energy_\elastindex = \int_\refmembrane d\reffl \left(
  \frac{\lamelambda+2\lamemu}{2}\left(\tr{\ten{\lagr}}\right)^2 + 2\lamemu
  \det{\ten{\lagr}} \right) \,,
\end{equation}
and the curvature term \citep{helfrich1973}
\begin{equation}
  \energy_\bendindex = \int\limits_\membrane
  d\fl\frac{\bend}{2}\left(2\mean-\spont\right)^2 = \int\limits_\refmembrane
  d\reffl \surdil \frac{\bend}{2}\left(2\mean-\spont\right)^2\,,
\end{equation}
for the bending energy. Here, $\bend$ is the bending rigidy and $\spont$ is
the spontaneous curvature, while $\lamelambda$ and $\lamemu$ are the
2d-Lam{\'e} coefficients in Hooke's law valid for small deformations. These
coefficients correspond to the surface shear modulus $G_s=\lamemu$ and the
surface Poisson ratio $\nu_s=\lamelambda/(\lamelambda+2\lamemu)$ in the
notation of \citet{barthes-biesel2002}. Other constitutive laws can trivially
be implemented.

\subsection*{Hydrodynamics}
For all experimental setups the Reynolds number is small and the flow is
governed by the linear Stokes equations. The velocity
$\vec{\vel}^\alpha(\vec{\ort})$ of the inner $(\alpha=\inindex)$ and outer
$(\alpha=\outindex)$ fluid at the position $\vec{\ort}$ is determined by the
incompressibility relation
\begin{equation}
   \div{\vec{\vel}^\alpha} = 0 \,,
   \label{eq:incompr}
\end{equation}
the linear momentum equation
\begin{equation}
  -\grad{\p^\alpha} + \visc^\alpha \laplace{\vec{\vel}^\alpha} = \vec{0}
  \label{eq:stokes}
\end{equation}
with the isotropic pressure $\p^\alpha$, and by appropriate boundary
conditions far away from the capsule and at the membrane.

The flow has to be regular everywhere and continuous across the membrane when
assuming a no-slip boundary condition.  The jump in hydrodynamic traction
between both fluids is compensated by the elastic forces at the membrane
\begin{equation}
  \vec{\force}(\lt,\lp) = \left[ \ten{\stress}^\inindex(\vec{\shape}(\lt,\lp)) -
    \ten{\stress}^\outindex(\vec{\shape}(\lt,\lp))\right]\cdot\vec{\normal}(\lt,\lp) \,,
  \label{eq:boundary_force}
\end{equation}
where $\ten{\stress}^\alpha$ is the inner and outer hydrodynamic stress tensor
with Cartesian components
\begin{equation}
  \stress_{ij}^\alpha \equiv - \delta_{ij} \p^\alpha + \visc^\alpha
  \left(\partial_i \vel_j^\alpha + \partial_j \vel_i^\alpha\right) \,.
\end{equation}
Since assuming no-slip boundary conditions, the velocity is continuous across
the membrane, and the membrane is advected with the flow
\begin{equation}
  \left.\vec{\vel}^\inindex(\vec{\ort})\right|_{\vec{\ort}=\vec{\shape}(\lt,\lp;t)} = \left.
  \vec{\vel}^\outindex(\vec{\ort})\right|_{\vec{\ort}=\vec{\shape}(\lt,\lp;t)}
  = \partial_t \vec{\shape}(\lt,\lp;t)\,. \label{eq:no_slip}
\end{equation}
Far away from the capsule the outer flow coincides with the
undisturbed external flow $\vec{\vel}^\infindex$
\begin{equation}
  \vec{\vel}^\outindex(\vec{\ort}) \to \vec{\vel}^\infindex(\vec{\ort})
  ~~~\text{for}~ \norm{\vec{\ort}} \to \infty \,.
  \label{eq:boundary_infinity}
\end{equation}
Since the Stokes equations are linear, we can split the total flow into a sum
of the undisturbed flow and an induced flow
\begin{equation}
  \vec{\vel}^\alpha \equiv \vec{\vel}^\infindex + \vec{\vel}^\alpha_\induced
  \,, \label{eq:split}
\end{equation}
where the homogeneous boundary condition
$\vec{\vel}^\alpha_\induced(\vec{\ort})\rightarrow 0$ far away from the
capsule is easy to implement.

For specific applications, we employ a linear shear flow (figure
\ref{fig:capsule2})
\begin{equation}
  \vec{\vel}^\infindex (\vec{\ort}) = \shearrate \, y \vec{\cart}_x
\end{equation}
with shear rate $\shearrate$. 
The equations of motion of the membrane are fully determined by Stokes'
equations (\ref{eq:incompr}, \ref{eq:stokes}), the force balance with the
elastic forces (\ref{eq:boundary_force}), and the boundary conditions
(\ref{eq:boundary_infinity}, \ref{eq:no_slip}) which include the membrane
advection (right hand side of \ref{eq:no_slip}).

\subsection*{Dimensionless parameters}
The motion of the capsule is governed by a number of dimensionless parameters:
The volume $\vol$ of the capsule remains constant and defines a length scale
$\lenscale$
\begin{equation}
  V \equiv \frac{4\pi}{3}\lenscale^3 \,,
\end{equation}
which will be used as the unit length. The elastic energy density is given by
the elastic moduli depending on the given constitutive law. In our case we use
the shear elasticity $\lamemu$ to define an energy scale $\lamemu
\lenscale^2$. The remaining elastic parameters can thus be cast in a
non-dimensional form by defining the two dimensional Poisson number
\begin{equation}
  \poisson \equiv \frac{\lamelambda}{\lamelambda+2\lamemu} \,.
\end{equation}
and the non-dimensional bending rigidy
\begin{equation}
  \dimbend \equiv \frac{\bend}{\lamemu \lenscale^2},\label{eq:1}
\end{equation}
and spontaneous curvature
\begin{equation}
    \dimspont \equiv \frac{\lenscale}{2} \spont \,.
 \end{equation}
The viscosity $\viscout$ can be used to define a time scale
$\lenscale\viscout/\lamemu$, giving the capillary number
\begin{equation}
  \dimshear \equiv \frac{\lenscale\viscout}{\lamemu} \shearrate \,.
\end{equation}
Finally, the viscosity contrast $\visccont$ has already been defined in
(\ref{eq:visc_cont}).

\section{Spectral Method} \label{sec:spectral}
We now develop a method to numerically solve the nonlinear equations
of motion.

\subsection*{Spectral method}
To transform the shape function to spectral space, we expand its Cartesian
components $\shape^i(\lt,\lp)\equiv \vec{\shape}(\lt,\lp)\cdot\vec{\cart}_i$
into scalar spherical harmonics $\y{l}{m}(\lt,\lp)$ with $l\ge0, \norm{m}\leq
l$ \citep{rose1957, brink1968}. More generally, we consider the spectral
expansion of a scalar function $f$
\begin{equation}
    f(\lt,\lp) = \sum\limits_{lm} f^{lm}\y{l}{m}(\lt,\lp)\,.
\end{equation}
Since the set of all spherical harmonics form a complete and orthonormal basis
on the sphere $S^2$:
\begin{equation}
  \int_{S^2} d\omega\, \cy{l}{m}(\lt,\lp)\y{l'}{m'}(\lt,\lp) = \delta_{mm'}\delta_{ll'}\,,
\end{equation}
the spectral coefficients are in principle obtained by the
integral
\begin{equation}
  f_{lm} = \int_{S^2}d\omega\, \cy{l}{m}(\lt,\lp) f(\lt,\lp) \,.
\end{equation}
Here $d\omega=\sin \lt d\lt d\lp$ denotes the area element on the sphere $S^2$
and the superscript star indicates the complex conjugate.  For numerical
applications, however, this integral must be replaced by a discrete sum. For
the hydrodynamic part it will be quite useful to extend the spherical
harmonics to $l<0$ by defining $Y^m_{-(l+1)} \equiv Y^m_l$.

A function whose spectral coefficients $f_{lm}$ vanish for $l\geq\bw$ is
called a bandlimited function with bandlimit $\bw$ {\citep{healy2003}}.  We
solve the dynamics of the membrane by restricting to the space of bandlimited
shape functions. Since the spectral amplitude of smooth functions decays
exponentially with $l$ {\citep{boyd2001}}, this scheme is very accurate
already for low $\bw$.

To get the expansion coefficients out of a given bandlimited function, we
choose a finite number of collocation markers $(\lt_i,\lp_i)$, $i=1,\ldots,n$
at the membrane. A scalar function $f(\lt,\lp)$ living on the membrane is then
determined by its values at these points $f_i\equiv f(\lt_i,\lp_i)$, whereas in
spectral space this function is represented by the spectral coefficients
$f^{lm}$ up to the bandwidth $\bw$. Transformation from spectral space to real
space is easily done by evaluating the spherical harmonics at the collocation
points
\begin{equation}
  f_i = \sum\limits_{lm} f^{lm} \y{l}{m}(\lt_i,\lp_i) \equiv \sum\limits_{lm}
  \mat_i^{lm} f^{lm} \,.
\end{equation}
If the transformation matrix $\mat_i^{lm}$ is square and regular, the inverse
transformation can be obtained by
\begin{eqnarray}
  f^{lm} = \sum\limits_i f_i \left(\tenpseudo\right)_i^{lm} \,,
\end{eqnarray}
where $\tenpseudo$ is the inverse of $\ten{\mat}$. {However,} often more
collocation points than spectral modes are used
{\citep{healy2003}}: A natural choice is to define the
collocation markers $i\equiv(j,k)$ equidistantly in $\lt$ and $\lp$, where we
shift the values $\lt$ away from the poles { to avoid numerical
  problems in the vicinity of the pole \citep{boyd2001}}
\begin{eqnarray}
  \lt_{(j,k)} &\equiv& \frac{(2j+1)\pi}{4\bw} \,, \\
  \lp_{(j,k)} &\equiv& \frac{k\pi}{\bw}
\end{eqnarray}
with $j,k=0,\ldots,2\bw -1$. With this choice the number of markers $n=4\bw^2$
is larger than the number of spectral modes $\bw^2$.  In this case
$\tenpseudo$ has to be replaced by the Moore-Penrose-Pseudoinverse
\citep{Swarztrauber2004}.

The expressions of the metric and curvature involve first order and second
order derivatives, respectively. The main advantage of spectral methods is
that the derivatives of the basis functions are known algebraically
\citep{rose1957, brink1968}, and therefore differentiation can be performed
with high accuracy for bandlimited functions. Similarly the integral over the
function $f$ is evaluated numerically via
\begin{equation}
  \int_{S^2}d\omega\,f(\lt,\lp) = \sqrt{4\pi}f_{00}\,,
\end{equation}
which follows easily from $\y{0}{0}(\lt,\lp)=1/\sqrt{4\pi}$.  Once the
derivatives of the shape functions are known, all further geometrical
computations are performed at the collocation points in physical space. It is
straightforward to numerically calculate the energy density of a given shape
for a given constitutive law with high accuracy. Similarly, the variation of
the energy density is evaluated for a given shape and variation of the shape
function.

\subsection*{Elastic forces}
To get the force density at the collocation points, we need to calculate the
variation of the total free energy for special variations of the shape. The
spectral coefficients of the force density are obtained in the most direct way
when we choose the specific variations
\begin{equation}\label{eq:shapevariation}
  \delta \vec{\shape}^i_{lm} (\lt, \lp) \equiv - \frac{\sin{\lt}}{\sqrt{\metric(\lt,
  \lp)}} \cy{l}{m}(\lt, \lp) \vec{\cart}_i \,,~~~l=0,\ldots,\bw\,;~m=-l,\ldots,l\,;
  ~i=x,y,z \,.
\end{equation}
The variation of the metric and curvature tensor resulting from this shape
variation can be evaluated easily using the known derivatives of the shape
function. Since the derivatives of the energy density with respect to metric
and curvature are known analytically, the variation of the elastic energy
(\ref{eq:energy_variation}) can be calculated easily.For the specific choice
of $\delta \vec{\shape}^i_{lm}$ (\ref{eq:shapevariation}), this yields
directly the Cartesian spectral force components
\begin{equation}
  \delta \energy = - \int_\membrane
  d\fl\, \vec{\force} \cdot \delta \vec{\shape}^i_{lm} = \int_{S^2}
  d\omega\, \vec{\force} \cdot \vec{\cart}_i \cy{l}{m}(\lt, \lp) =
  \force^i_{lm} \,.
\end{equation}

\subsection*{Hydrodynamics}
We follow a similar strategy for the hydrodynamic part of the problem. To
solve the hydrodynamic equations, we choose a complete set of basis functions
in three dimensional space that automatically fulfill Stokes' equations. These
so called Lamb modes \citep{lamb1932, happel1983} can be defined as
appropriate linear combinations of vector spherical harmonics.  Spherical
coordinates $\sr$, $\sth$ and $\sph$ in the laboratory frame as well as the
corresponding basis vectors $\vec{\spherical}_\sr$, $\vec{\spherical}_\sth$
and $\vec{\spherical}_\sph$ are best suited for calculations concerning the
Lamb modes. We stress again the difference between the Lagrangian coordinates
$(\lt,\lp)$, which serve as markers for the material points, and the angles
$(\sth,\sph)$, which are spherical coordinates in the laboratory frame. With
help of the surface gradient on the sphere $S^2$
\begin{equation}
  \surgrad \equiv \vec{\spherical}_\sth \partial_\sth +
  \frac{1}{\sin{\sth}}\vec{\spherical}_\sph \partial_\sph
\end{equation}
the vector spherical harmonics are given by \citep{morse1953}
\begin{eqnarray}
  \vy{l}{m}(\sth,\sph) &\equiv& \y{l}{m}(\sth,\sph) \vec{\spherical}_\sr\,,\\
  \vpsi{l}{m}(\sth,\sph) &\equiv& \frac{1}{\sqrt{l(l+1)}} 
     \surgrad \y{l}{m}(\sth,\sph) \,,\\
  \vphi{l}{m}(\sth,\sph) &\equiv& \frac{1}{\sqrt{l(l+1)}} 
     \vec{\spherical}_\sr \times \surgrad \y{l}{m}(\sth,\sph) \,.
\end{eqnarray}
The Lamb modes then read
\begin{eqnarray}
  \lambp{l}{m}(\sth,\sph) &\equiv& 
          \frac{1}{2(l+1)(2l+3)}\left[l(l+1)\vy{l}{m}(\sth,\sph)+(l+3)
          \sqrt{l(l+1)}\vpsi{l}{m}(\sth,\sph)\right] \,,\\
  \lambphi{l}{m}(\sth,\sph) &\equiv&  
          l\vy{l}{m}(\sth,\sph)+\sqrt{l(l+1)}\vpsi{l}{m}(\sth,\sph) \,,\\
  \lambchi{l}{m}(\sth,\sph) &\equiv&  -\sqrt{l(l+1)}\vphi{l}{m}(\sth,\sph) \,.
\end{eqnarray}
With the scalar spherical harmonics and the Lamb modes as basis functions the
pressure and velocity fields can be expanded for the external and internal as
well as the induced flow
\begin{eqnarray}
  \p^\alpha (\sr,\sth,\sph) &=& \visc^\alpha \sum_{lm} p^\alpha_{l,m} \sr^l  
         \y{l}{m}(\sth,\sph) \,,\\
  \vec\vel^\alpha (\sr,\sth,\sph) &=& \sum_{lm} \left( p^\alpha_{l,m}
    \sr^{l+1}  \lambp{l}{m}(\sth,\sph) + \phi^\alpha_{l,m}
    \sr^{l-1}  \lambphi{l}{m}(\sth,\sph) + \chi^\alpha_{l,m}
    \sr^{l}  \lambchi{l}{m}(\sth,\sph)  \right)\,.
\end{eqnarray}
Due to the regularity of the induced flow at the origin and the boundary
conditions at infinity (\ref{eq:boundary_infinity}), the sums are restricted
to $l\ge 0$ for $\vec{\vel}^\inindex_\induced$ and $l\le -2$ for
$\vec{\vel}^\outindex_\induced$.

The remaining boundary conditions (\ref{eq:no_slip}) and force balance
condition (\ref{eq:boundary_force}) result in linear equations for the
coefficients of the Lamb modes. For our choice of collocation points, we have
an overdetermined system that can be solved in a least square sense. This way
we get the hydrodynamic flow for any given deformation of the capsule.

Since the capsule is simply advected with the flow (\ref{eq:no_slip}), we
perform Euler steps with a sufficently small time step to determine the
dynamics of the capsule. Higher order methods such as a second order
Runge-Kutta have been tested, but did not yield significant improvements in
terms of stability and overall simulation time.

\section{Spherical Capsules} \label{sec:spherical} To test our method, we
compare with quasispherical results that can be obtained analytically. These
tests comprise the relaxation dynamics of a capsule to its spherical reference
shape in quiescent fluid and the stationary deformation of a capsule in linear
shear flow for low shear rates.

Small deformations of an initially spherical capsule relax exponentially in
time. \citet{rochal2005} have identified the normal modes, and calculated the
corresponding relaxation times. The relaxation modes are linear combinations
of vector spherical harmonics and are likewise labeled by $l,m$. For each set
of angular momentum numbers, three relaxation normal modes exist, which have
been termed ``stretching'', ``bending'', and ``shear'' mode, respectively. The
corresponding relaxation times are obtained from the eigenvalue equation (24)
of \citet{rochal2005}.

\begin{figure}
  \begin{center}
    \includegraphics[width=0.79\linewidth]{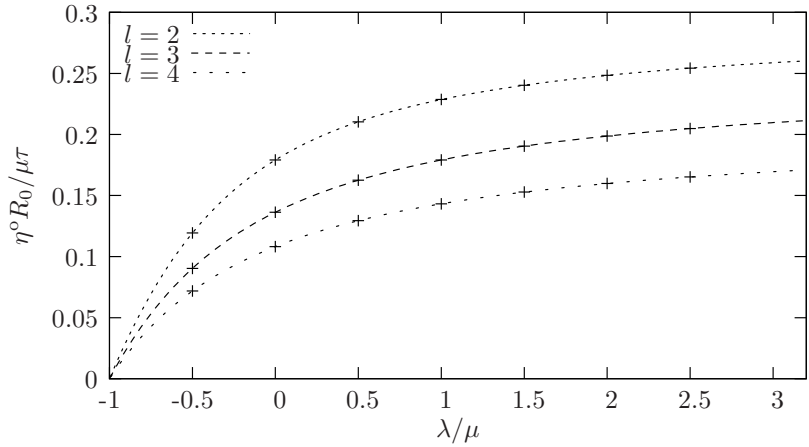}
    \caption{Comparison of numerically and analytically obtained relaxation
      time. --- The plot shows numerically obtained values of the scaled
      inverse relaxation time $\viscout\lenscale/\lamemu\tau$ of bending modes
      as a function of the ratio $\lamelambda/\lamemu$ for different harmonic
      indexes $l=2$, $l=3$, $l=4$ and $\bend=0$. The curves are the analytic
      solution of the secular equation (24) of \citet{rochal2005}.}
    \label{fig:relaxation-bending}
  \end{center}
\end{figure}

As a numerical test of our code, a spherical capsule was deformed in the
direction of a normal mode, and the time constant of the subsequent relaxation
to equilibrium was extracted. In figure \ref{fig:relaxation-bending}, we
compare the relaxation times of selected modes as a function of the Lam{\'e}
parameter $\lambda$ with the theoretical predictions, with excellent
agreement.

Switching on the shear flow with a low shear rate, the capsule relaxes into a
stationary shape with a tank-treading motion. The stationary deformations have
been calculated to first and second order in the deformation by
\citet{barthes-biesel1981}. The deformation of the capsule is measured by the
time-dependent Taylor deformation parameter
\begin{equation}
  \deform \equiv \frac{\longax-\shortax}{\longax+\shortax} \,,
  \label{eq:deformdefinition}
\end{equation}
where $\longax$ is the longest and $\shortax$ is the shortest distance of the
membrane from the center (see figure \ref{fig:capsule2}). In the long time
limit, $\deform$ assumes a stationary value $\deform_0$, which is to first order in
$\dimshear$
 ($\dimspont=1$) \begin{equation}
\label{eq:deformqs}  \deform_0 \approx 
  \frac{5}{4}\frac{\poisson+2}{\poisson+1+2\dimbend(\poisson+5)}\dimshear 
  = \frac{5}{8}
  \frac{(4\lamemu+3\lamelambda)\lenscale\viscout}{(\lamemu+\lamelambda)
   \lamemu+\frac{2\bend}{\lenscale^2}(3\lamelambda+5\lamemu)}\shear
  \,.
\end{equation}

In our simulations, the unstressed capsule is subjected to an abruptly started
linear shear flow. Numerically, the deformation and inclination are not
calculated directly according to definition~(\ref{eq:deformdefinition}). As
described by \citet{ramanujan1998}, we find it more suitable to use instead
the deformation parameter of an ellipsoid with the same tensor of inertia. In
the small deformation limit, both definitions are equivalent.

\begin{figure}
 \begin{center}
  \begin{minipage}{0.49\linewidth}
    (a)\\
    \includegraphics[width=0.99\linewidth]{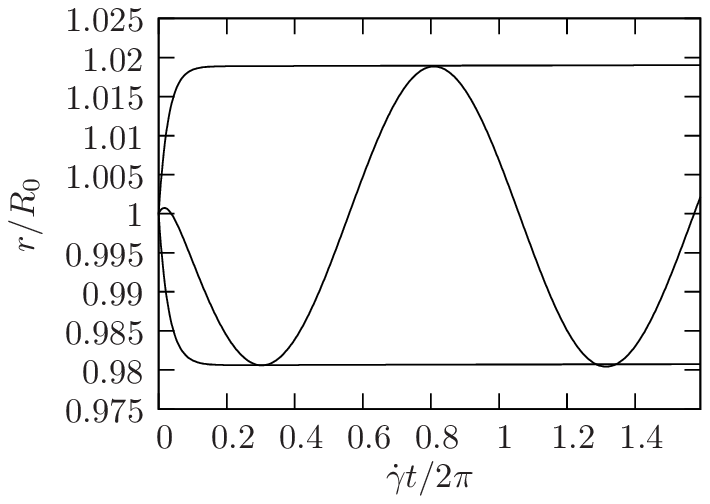}
  \end{minipage}
  \begin{minipage}{0.49\linewidth}
    (b)\\
    \includegraphics[width=0.99\linewidth]{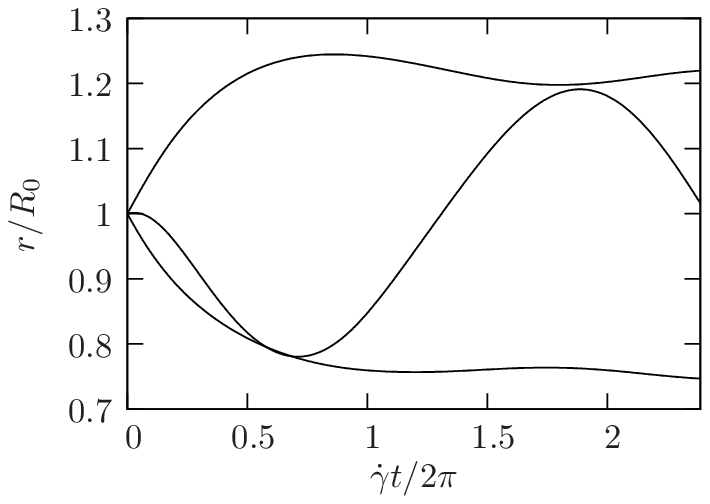}
  \end{minipage}
  \begin{minipage}{0.49\linewidth}
    (c)\\
    \includegraphics[width=0.99\linewidth]{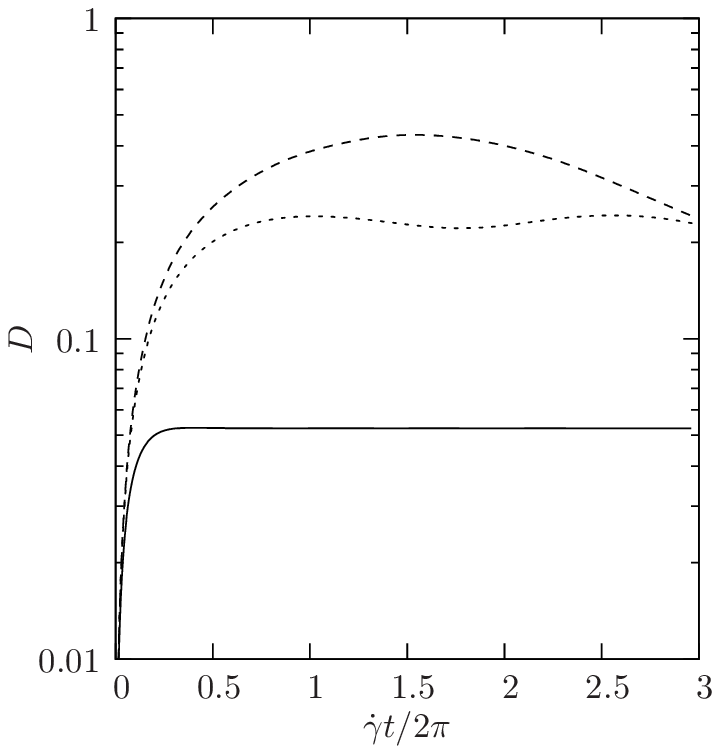}
  \end{minipage}
  \begin{minipage}{0.49\linewidth}
    (d)\\
    \includegraphics[width=0.99\linewidth]{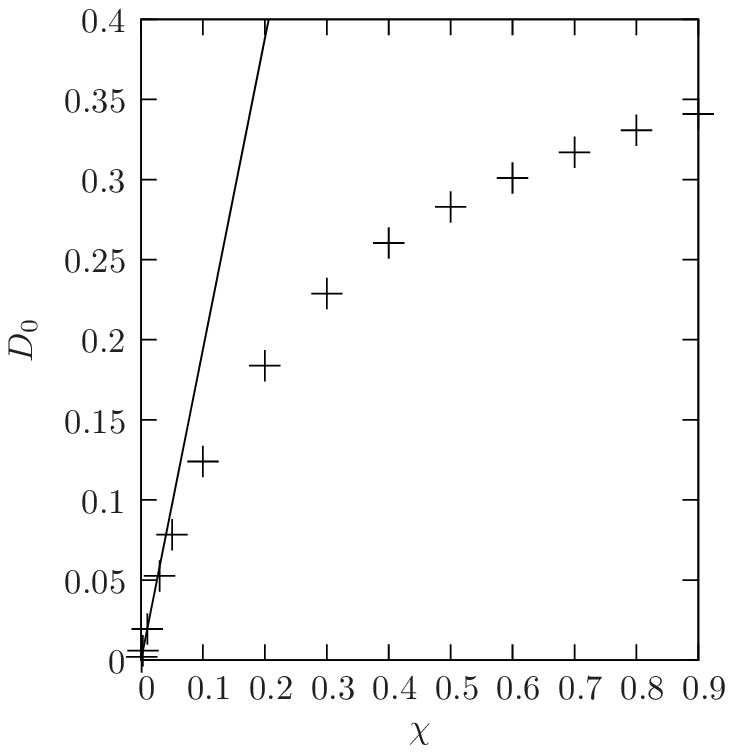}
  \end{minipage}
  \caption{Dynamics of an initially spherical capsule in shear flow. --- (a)
    and (b) Time evolution of maximal, minimal radius and radius of a tracer
    particle for abruptly starting shear flow at time $t=0$ for different
    shear rates $\dimshear=0.01$ (a), $\dimshear=0.3$ (b) --- (c) Time
    evolution of deformation parameter $\deform$ for different shear rates:
    $\dimshear=0.03$ (continuous), $\dimshear=0.3$ (dotted), $\dimshear=3.$
    (dashed line). --- (d) Plot of the stationary deformation parameter
    $\deform_0$ as a function of the dimensionless shear rate $\dimshear$. For
    low shear rates the results of the linear theory (eq. \ref{eq:deformqs},
    full line) are approached whereas for higher shear rates deviations are
    clearly visible. --- Constant parameters for (a)-(d): viscosity contrast
    $\visccont=10$, Poisson number $\poisson=0.5$, non-dimensional bending
    rigidy $\dimbend=0.01$ and spontaneous curvature $\dimspont=1.$}
  \label{fig:deform_evolution}
 \end{center}
\end{figure}

The simulations were performed by using a bandwidth of $b=11$ corresponding to
$121$ modes, and a total of $N=4\bw^2=484$ grid points. The bending energy was
chosen to prevent the formation of wrinkling modes within the bandwidth. The
time step was chosen small enough to yield a stable evolution. Typical values
were $\Delta t \sim 1 / (1000 \shearrate)$. The time evolution of the
deformation parameter of an initially spherical capsule is shown in figure
\ref{fig:deform_evolution} for different shear rates and fixed elastic
parameters. To illustrate the stationary tank-treading motion of the capsule,
we also show the distance of a marker point on the membrane to the center of
the capsule as a function of time. This distance oscillates with twice the
tank-treading frequency between $\longax$ and $\shortax$. At low shear rate,
we observe a monotonic relaxation of the deformation to its stationary value
$\deform_0$, while for large $\dimshear$ and $\visccont$ a pronounced
over-shoot is observed \citep[c.f.][]{ramanujan1998}. 
The numerical deformation $\deform_0$ clearly follows the asymptotic
prediction~(\ref{eq:deformqs}) for low shear rate. For large shear rates, the
simulation data show deviations from linear behaviour. These are likely due to
the rotational part of the linear shear flow, since for non-zero vorticity the
stationary deformation is a non-linear function of the shear rate even in the
quasispherical limit.

\section{Ellipsoidal Capsules} \label{sec:ellipsoidal}
After successfully testing our spectral method by means of a spherical initial
or reference shape, we can continue to investigate the dynamics of capsules
with an ellipsoidal initial shape. This case is both experimentally more
realistic, since synthesised capsules are never perfectly spherical, and leads
to a richer dynamical behaviour of the capsules. In a non-spherical reference
shape, the membrane points are no longer equivalent to each other. During the
course of a tank-treading motion, a membrane element is therefore periodically
deformed, which costs elastic energy. Any membrane element therefore
energetically prefers its initial position (or one of the equivalent positions
by symmetry of the reference shape). This effect is called ``shape memory'',
and plays an important role also in the dynamics of red blood cells
\citep{fischer2004,takatani2006}.

The shape memory effect is the cause of oscillations of the deformation and
the inclination angle in the tank-treading state, as observed experimentally
by \citet{chang1993}, \citet{walter2001}, and \citet{abkarian2007}, and also
found in the simulations by \citet{ramanujan1998}. Recently, a modified
Keller-Skalak type theory was proposed \citep{skotheim2007}, which explains
this behaviour qualitatively. Their model also predicts an oscillating
tank-treading motion at large shear rate, and a tumbling motion at lower shear
rate. In the tumbling regime, a tracer particle on the membrane oscillates
around a fixed position with respect to the capsule shape. In the intermediate
shear rate regime, intermittent motion, which alternates between tumbling and
tank-treading, is predicted. Although direct experimental evidence for this
behaviour is missing, indirect evidence was provided by \citet{abkarian2007},
who discovered a hysteresis of the transition from tumbling to tank-treading
and the reverse transition by increasing or decreasing the shear rate,
respectively. We use our spectral method to systematically explore the full
phase diagram in a large range in shear rate as well as viscosity contrast.
Thus the quantitative accuracy of the reduced model in \citet{skotheim2007}
can be tested. While \citet{ramanujan1998} observed the onset of a tumbling
motion for low sphericity, due to the formation of cusp-like instabilities in
the shape the simulations never went beyond half a tumbling motion. Grid
distortion also required the use of explicit numerical smoothing in more
recent simulations \citep{pozrikidis2003a}. Since bending rigidity is included
in our method, the formation of cusps can be suppressed, leading to a more
stable algorithm. The cutoff at a finite bandwidth in our algorithm also
effectively implements numerical smoothing. Nevertheless high order modes
accumulate numerical errors during the simulation run in particular at large
shear rates, thereby limiting the maximum simulation time.

\subsection*{Phase Diagram}
\begin{figure}
 \begin{center}
  \includegraphics{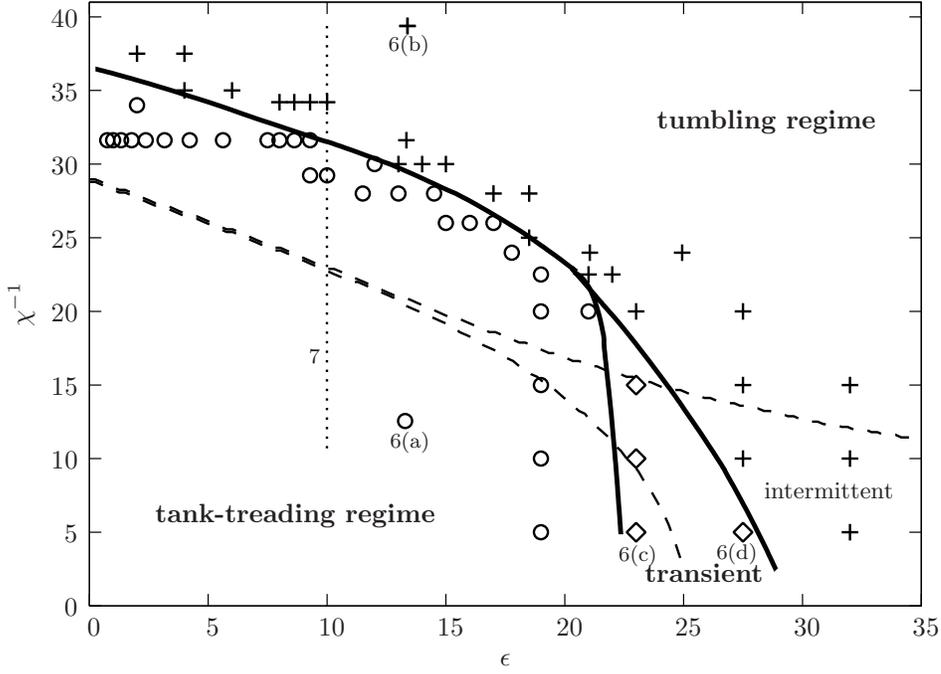}
  \caption{Phase diagram of an elastic capsule in shear flow with the tumbling
    and tank-treading regimes as a function of the viscosity contrast
    $\visccont$ and the inverse dimensionless shear rate $\dimshear^{-1}$. ---
    The full line is a guide to the eye separating the tank-treading
    (circles), tumbling (crosses) and transient region (diamonds) for our
    simulation. Dashed lines indicate the phase diagram due to
    \citet{skotheim2007} for the same parameter set. In the region between the
    dashed lines intermittent motion is predicted. We have not found conclusive
    evidence for this kind of motion, but rather found transient dynamics from
    tumbling to tank-treading. The numbers correspond to following figures,
    where parts of the phase diagram are examined closer. Geometrical
    parameters: $a_2=a_3=0.9a_1$, elastic parameters: $\poisson=0.333$,
    $\dimbend=0.01$, $\dimspont=1.$}
  \label{fig:phasediagram}
 \end{center}
\end{figure}

Our numerical results for the overall phase diagram are summarised in figure
\ref{fig:phasediagram}, where the dynamical behaviour is plotted as a function
of the inverse dimensionless shear rate $\dimshear^{-1}$ and the viscosity
contrast $\visccont$. At low shear rate, the hydrodynamic forces are too small
to overcome the energy barrier present for a tank-treading motion due to the
shape memory effect. Therefore, capsules tumble at low $\dimshear$, while an
oscillating tank-treading behaviour is stable at large $\dimshear$. We also
observe a transient dynamics from tumbling to tank-treading for large
viscosity contrast $\visccont$, which will be discussed below.  Although this
transient dynamics might be taken as indications of intermittent motion, we
could not find conclusive evidence during the time of our simulation runs. In
particular, we never observed a transition from tank-treading to tumbling.
Also shown in this figure is the phase diagram for the analytic model by
\cite{skotheim2007}. The qualitative agreement, apart from the apparent lack
of intermittent behaviour, seems to be rather good, given the crude dynamics
implemented in the reduced analytical model. Only at large viscosity contrast
pronounced differences in the shape of the phase diagram start to feature.
Closer inspection of the data reveals significant oscillations of the axis
lengths, which are fixed in the reduced model.  These breathing modes may
alter the intermittent character of the capsule motion in the model by
\cite{skotheim2007}.

For all simulations the deformation remained well within the range of validity
of Hooke's law. The extension ratios (c.f.~appendix \ref{app_dg}) never
deviated more than 5\% from unity. The results are therefore not susceptible
to the specific elastic law in this regime.

\subsection*{Oscillation Amplitudes}
We proceed to quantify the oscillations in the tank-treading and tumbling
state and investigate the transient dynamics below. For the definition of
inclination $\incl$ and tank-treading angle $\tank$ see figure
\ref{fig:capsule2}. They are defined as the angles between the flow direction
and the maximal radius or a marker point, respectively.  Initially, these
angles are chosen to lie in the interval $[0,\pi[$. To make both $\tank(t)$
and $\incl(t)$ continuous functions of time, we manually add $2\pi$ after a
full rotation, thereby keeping the variation of each angle between subsequent
time steps as small as possible.  As a consequence, the angles can assume
values outside the interval $[0,2\pi[$ at later times. This allows to count
the number of full rotations, which would not be possible if all angles were
restricted to $[0,2\pi[$. The phase shift of a material point away from its
elastic minimum
\begin{equation}
   \phase(t) \equiv (\tank(t) - \incl(t)) - (\tank(0) - \incl(0))
\end{equation}
is called phase angle.

\begin{figure}
  \begin{center}
    \begin{minipage}{0.99\linewidth}
      (a)\\
      \includegraphics[width=0.99\linewidth]{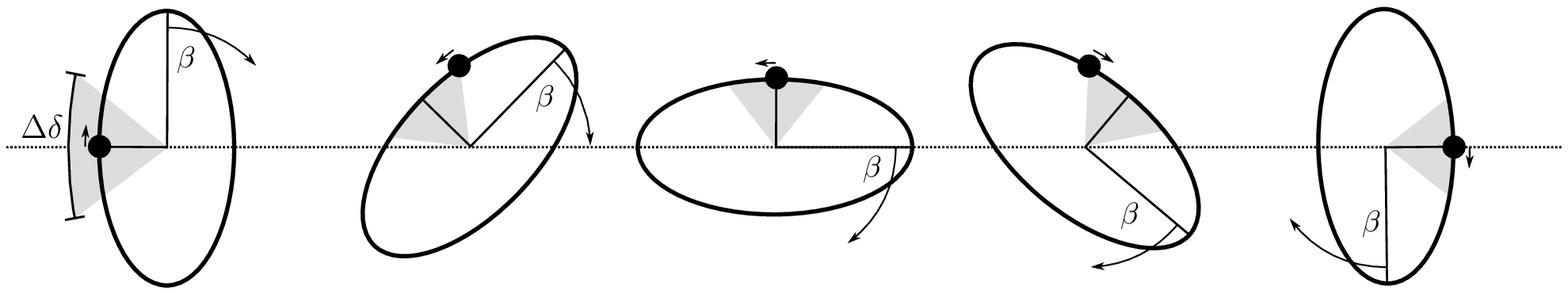}
    \end{minipage}
    \begin{minipage}{0.99\linewidth}
      (b)\\
      \includegraphics[width=0.99\linewidth]{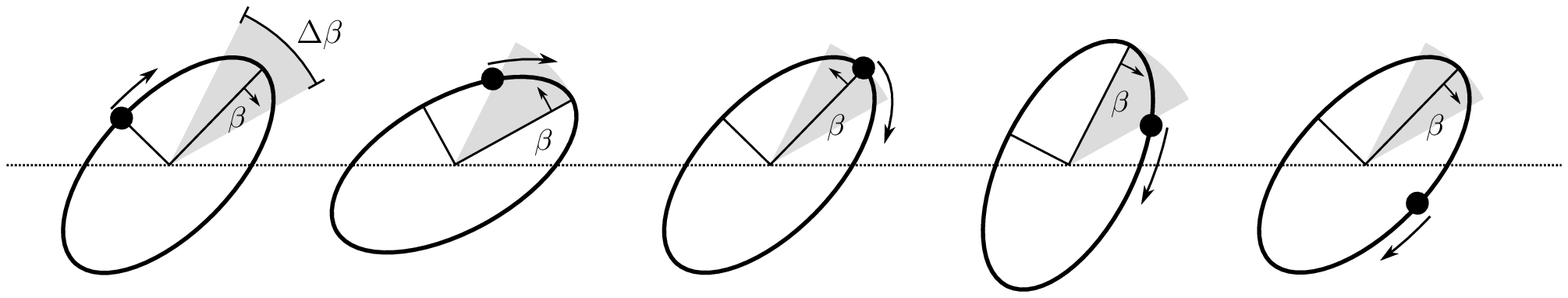}
    \end{minipage}
  \end{center}
  \caption{Definition of oscillation amplitudes of phase and inclination angle
    in tumbling and tank-treading states of motion. --- (a) shows the tumbling
    state: Here the inclination angle $\incl$ changes monotonously while the
    phase angle $\phase$ oscillates with amplitude $\Delta\phase$
    {as is indicated by a tracer particle.} --- (b) shows the
    tank-treading motion, where the inclination angle $\incl$ oscillates,
    while the phase angle $\phase$ and the tank-treading angle $\tank$ change
    monotonously.}
  \label{fig:oscis}
\end{figure}

\begin{figure}
 \begin{center}
\begin{minipage}{0.49\linewidth}
  (a)\\
  \includegraphics[width=0.99\linewidth]{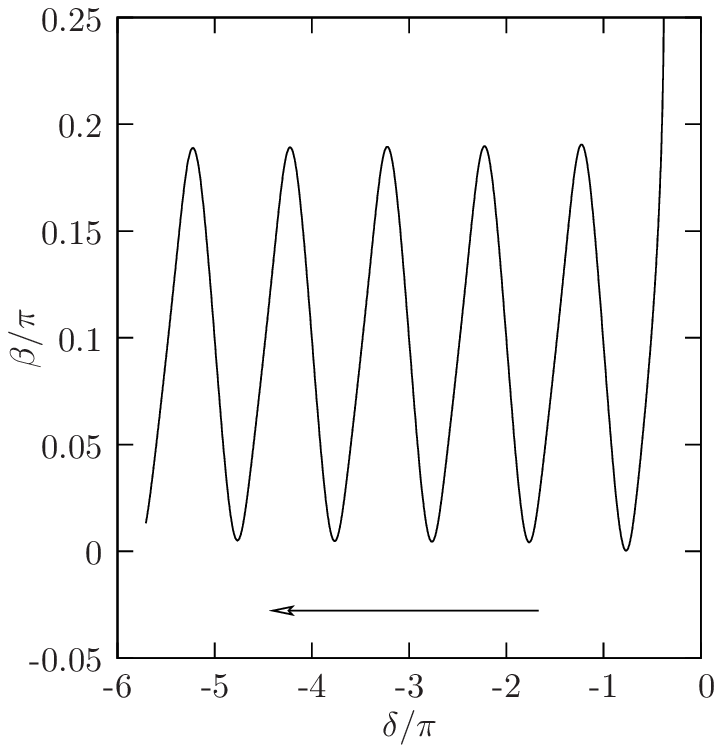}
\end{minipage}
\begin{minipage}{0.49\linewidth}
  (b)\\
  \includegraphics[width=0.99\linewidth]{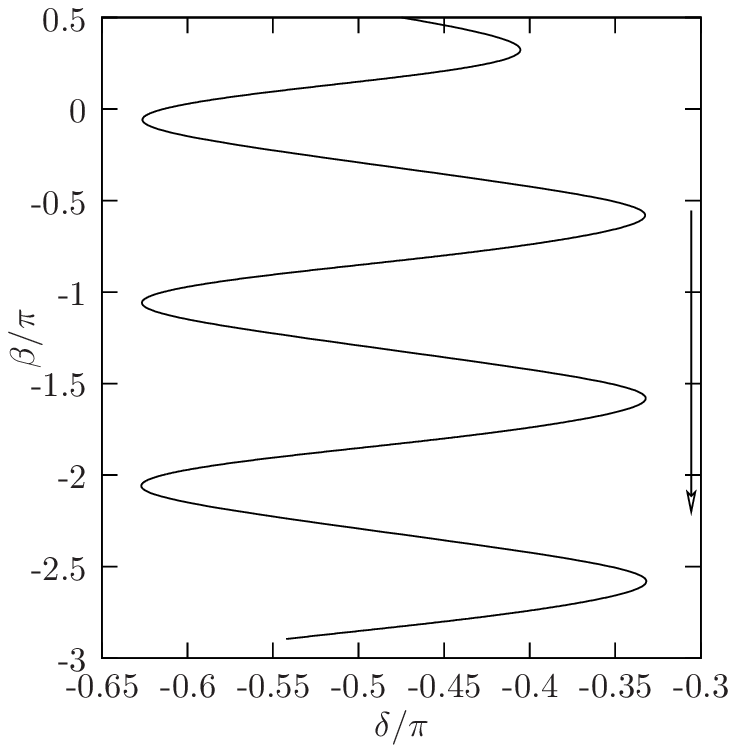}
\end{minipage}
\begin{minipage}{0.49\linewidth}
  (c)\\
  \includegraphics[width=0.99\linewidth]{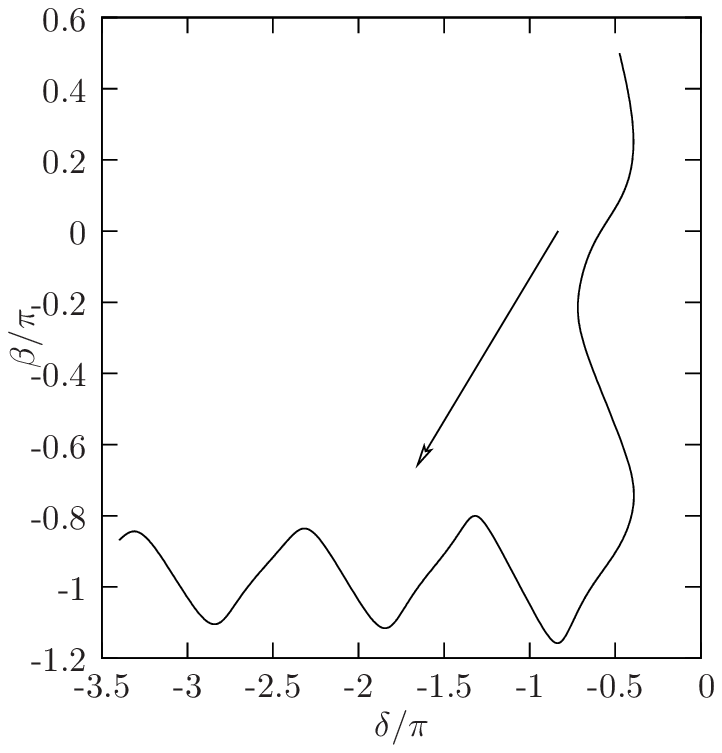}
\end{minipage}
\begin{minipage}{0.49\linewidth}
  (d)\\
  \includegraphics[width=0.99\linewidth]{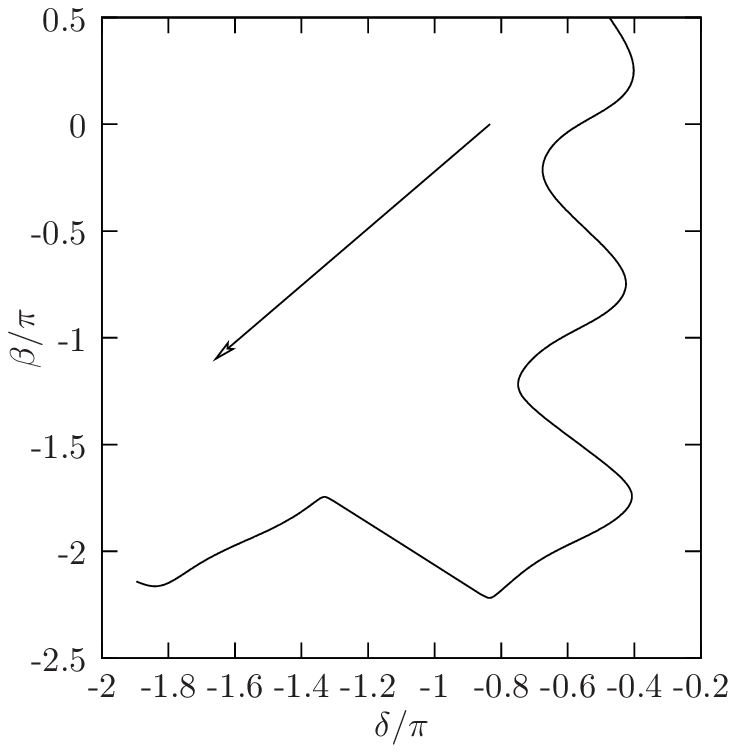}
\end{minipage}
\caption{Typical plots of inclination angle $\incl$ vs. phase angle $\phase$
  for tank-treading, tumbling and transient dynamics. --- The site of these
  plots in the phase diagram are labelled in figure \ref{fig:phasediagram} by
  corresponding numbers. --- (a) Typical tank-treading motion: $\incl$
  oscillates around a stable value, while $\phase$ changes monotonously. The
  arrow denotes the direction in time, viscosity contrast $\visccont=13.3$,
  non-dimensional shear rate $\dimshear=0.08$. --- (b) Typical tumbling
  motion: $\phase$ oscillates around a stable value, while $\incl$ changes
  monotonously, $\visccont=13.3$, $\dimshear=0.025$.  --- (c), (d) Typical
  motions for tumbling to tank-treading transition, $\visccont=23$,
  $\dimshear=0.2$ in (c) and $\visccont=27.5$, $\dimshear=0.2$ in (d).  ---
  The remaining parameters are equal to those used in figure
  \ref{fig:phasediagram}.}
  \label{fig:three_states}
 \end{center}
\end{figure}

In the tank-treading regime the inclination angle $\incl$ oscillates around a
stationary value $\incl_0$ with amplitude $\Delta\incl$ while the tank-treading angle
$\tank$ or phase angle $\phase$ changes monoto\-nously with time (see figure
\ref{fig:oscis}). In the tumbling regime, the inclination angle $\incl$
changes monoto\-nously with time, while the phase angle $\phase$ oscillates
around a fixed value $\phase_0$ with an oscillation amplitude $\Delta \phase$. Figure
\ref{fig:three_states} shows parametric plots of inclination vs. phase angle
for a tumbling and a tank-treading motion, as well as a transition from
tumbling to tank-treading.  The arrows indicates the direction of time in this
plot. In Figure \ref{fig:three_states} (c) and (d), one can see the transition
from a tumbling motion to an oscillating tank-treading motion near the
transition. Despite intensive search, we have not observed the reverse
behaviour: A tank-treading capsule never started to tumble. This might
indicate that the initial tumbling motion is only transient.

\begin{figure}
 \begin{center}
  \includegraphics{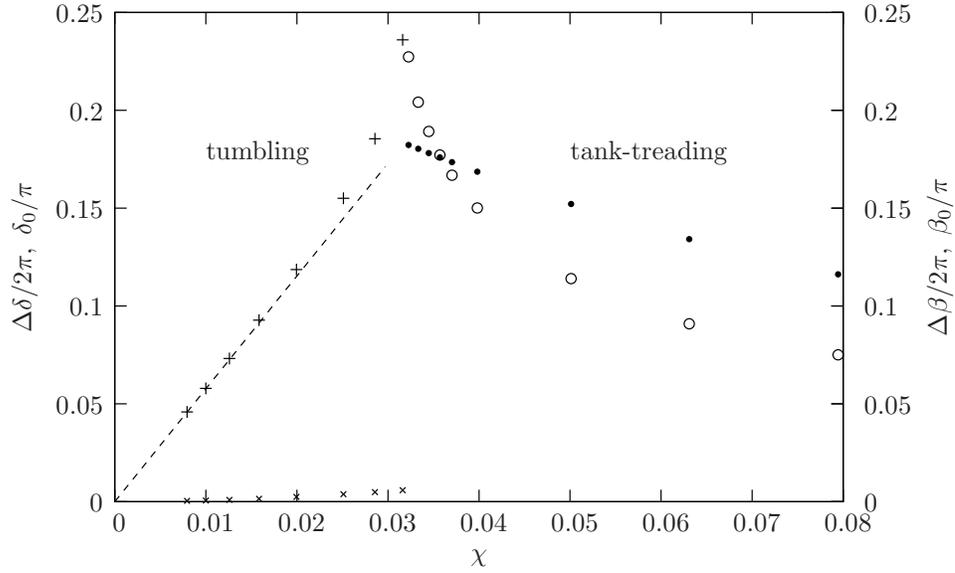}
  \caption{{Mean phase angle $\phase_0$ (small crosses), mean
      inclination angle $\incl_0$ (small filled circles) and oscillation
      amplitudes of phase angle $\Delta\phase$ (plus) and inclination angle
      $\Delta\incl$ (circles)} for different shear rates $\dimshear$ and a
    constant viscosity contrast $\visccont=10$. --- This cut through the phase
    diagram with $\visccont=10$ is indicated by the dotted line in figure
    \ref{fig:phasediagram}. At low shear rates the capsule tumbles with
    $\Delta\phase<\pi/2$ where the dashed line indicates a linear behaviour
    for small $\dimshear$, at higher shear rates the capsule tank-treads with
    $\Delta\incl<\pi/2$. --- remaining parameters as in figure
    \ref{fig:phasediagram}.}
  \label{fig:osci_shear}
 \end{center}
\end{figure}

For a fixed viscosity contrast $\visccont=10$ figure \ref{fig:osci_shear}
shows the shear rate dependence of the mean inclination angle $\incl_0$ in the
tank-treading regime, of the mean phase angle $\phase_0$ in the tumbling
regime, and of the oscillation amplitudes in both the tumbling and the
tank-treading regime. For low shear rates, in the tumbling regime, the mean
phase angle $\phase_0$ and the oscillation amplitude $\Delta \phase$ of the
phase angle are plotted as a function of the shear rate. Whereas the mean
phase angle depends only weakly on the shear rate, the oscillation amplitude
of the phase angle increases for increasing shear rate. For low shear rates,
this amplitude grows approximately linearly with the shear rate. When the
amplitude reaches approximately $\pi/2$, the capsule starts to tank-tread.

For higher shear rates, in the tank-treading regime, the mean inclination
angle $\incl_0$ and the oscillation amplitude $\Delta \incl$ of the
inclination angle are plotted as a function of $\dimshear$. With decreasing
shear rate, the oscillation amplitude of the inclination angle increases until
it reaches approximately $\pi/2$, where the transition to tumbling takes
place. The mean inclination angle also increases, but does not reach $\pi/4$.
Perhaps surprisingly the oscillation amplitude can be larger than the mean
inclination angle $\incl_0$ in the tank-treading regime, implying that the
inclination angle is negative for short periods of time.

\section{Summary} \label{sec:discussion}
During the last few years, the dynamics of elastic capsules in linear shear
flow has received increasing attention. Theoretical descriptions restricting
the capsule deformation to a few degrees of freedom predicted a rich dynamical
phase diagram, comprising of tank-treading, tumbling and an intermittent
motion. While recent experiments have found a hysteresis in the tank-treading
to tumbling transition for varying shear rate, direct observations of
intermittent behaviour is lacking so far.

To investigate elastic capsule systems, while maintaining complete control
over the underlying equations of motion, we implemented a spectral method to
numerically solve the equations of motion for a capsule. The capsule
deformation is expanded into smooth basis functions, leading to accurate
estimates on the membrane forces. The code is flexible and stable enough to
permit simulations for a large range of shear rates and viscosity contrasts
between inner and outer fluids.

Using this code, we could quantitatively recover the asymptotically known
stationary deformations of initially spherical capsules for low shear rate.

Finally we applied the numerical method to the ellipsoidal capsule system
similar to the one discussed by \citet{skotheim2007}. We observe a stable
tank-treading motion for large shear rate, in which the inclination angle
oscillates with twice the tank-treading frequency. At lower shear rate, or
higher viscosity contrast, the capsule starts to tumble. We systematically
explored the capsule dynamics over a wide range of viscosity ratios and shear
rates. The resulting phase diagram is qualitatively similar to the theoretical
predictions made by \citet{skotheim2007}, with the exception of intermittent
dynamics: While dynamic transitions from tumbling to a stable tank-treading
motion were observed, the reverse transition could not be observed.  An
analysis of the results suggests that the tumbling motion is only transient in
this part of the phase diagram.

Much longer simulation runs and a more detailed analysis near the presumed
intermittency region are needed to decide whether intermittent motion is
merely an artifact of the reduced dynamics employed by \citet{skotheim2007}.
Differences from the phase diagram at large viscosity ratios are likely due to
the restriction to a fixed capsule shape in the reduced model.

In conclusion, the spectral method developed in this work is a stable and
accurate complement to existing numerical methods. It allows the systematic
exploration of capsule dynamics over a wide range of material constants.
Theoretical predictions of the phase diagram of ellipsoidal capsules are
qualitatively confirmed, although quantitative differences exist, especially
for large viscosity ratios. The nature of predicted intermittent behaviour
warrants further investigation.

\begin{acknowledgements}
  Financial support of the DFG with in the priority programme SPP 1164 ``Nano-
  and Microfluidics'' is gratefully acknowledged.
\end{acknowledgements}

\oneappendix
\section{Differential geometry for deformed capsules} \label{app_dg} We first
recall some definitions of differential geometry that can be found in
\citet{frankel1997} and \citet{marsden1983}. We cover the two dimensional
surface with a coordinate net as outlined in section \ref{sec:problem}, where
the coordinates $(\lt,\lp)$ label the material points. The location of the
surface at time $t$ is given by the shape function $\vec\shape(\lt,\lp;t)$.
The basis vectors, defined by
\begin{equation}
   \vec{\basis}_i \equiv \partial_i \vec{\shape}\,,~~ i=\lt,\lp \,,
   \label{eq:basis}
\end{equation}
span the tangent planes and induce the outward pointing normal unit vector
\begin{equation}
   \vec{\normal} \equiv \frac{1}{\norm{\vec{\basis}_\lt \times
       \vec{\basis}_\lp}}\vec{\basis}_\lt \times \vec{\basis}_\lp \,.
   \label{eq:normal}
\end{equation}
They also define the metric tensor $\ten{\metric}$ with covariant components
\begin{equation}
   \metric_{ij} \equiv \vec{\basis}_i \cdot \vec{\basis}_j \,, 
   \label{eq:metric}
\end{equation}
where we use the ordinary three dimensional Euclidian scalar product. The
inverse metric tensor $\ten{\metric}^{-1}$ with contravariant components is
given by
\begin{equation}
   \metric^{ij}\metric_{jk}\equiv\kronecker^i{}_k \,, 
   \label{eq:inverse_metric}
\end{equation}
where Einstein's sum convention is employed. The determinant of the metric
tensor
\begin{equation}
   \metric \equiv \det{\ten{\metric}}
   \label{eq:det_metric}
\end{equation}
is connected to the basis vectors via
\begin{equation}
  \sqrt\metric=\norm{\vec{\basis}_\lt \times \vec{\basis}_\lp}\,.
\end{equation}

The area of an infinitesimal patch with width $d\lt$ and length $d\lp$ in
Lagrange coordinates is given by
\begin{equation}
   d\fl = \sqrt{\metric}d\lt d\lp \,.
   \label{eq:area}
\end{equation}
The curvature tensor $\ten{\curv}$, defined by
\begin{equation}
   \curv_{ij} \equiv \vec{\basis}_i \cdot \partial_j \vec{\normal} = - \vec{\normal}
   \cdot \partial_i \vec{\basis}_j \,,
   \label{eq:curvature}
\end{equation}
measures how the normal vector $\vec{\normal}$ changes its direction, when one
moves along the membrane. Its eigenvalues $k_{1,2}$ are called principle
curvatures and are the inverse radii of the principle curvature circles.
Trace and determinant of the curvature tensor define mean $\mean$ and Gaussian
curvature $\gauss$ respectively. They serve as scalar invariants of the
curvature tensor and can be expressed by the sum and the product of the
principle curvatures
\begin{equation}
   2\mean \equiv \tr{\ten{\curv}} = g^{ij}k_{ji} = \curv_1 + \curv_2 \,,~~
   \gauss \equiv \det{\ten{\curv}} = \curv_1\curv_2 \,.
   \label{eq:mean}
\end{equation}

In order to describe a deformation and an elastic response, an unstressed
reference shape $\vec\refshape(\lt,\lp)$ has to be specified as was mentioned
in section \ref{sec:problem}. The corresponding basis vectors
$\vec{\refbasis}$, normal vector $\vec{\refnormal}$ and metric
$\ten{\refmetric}$ are defined analogously.

The Lagrangian strain tensor $\ten{\lagr}$ with covariant components
\begin{equation}
   \lagr_{ij} \equiv \frac{1}{2}\left(\metric_{ij}-\refmetric_{ij}\right) \,,~~ i=\lt,\lp
   \label{eq:lagr}
\end{equation}
measures the change in length elements of the membrane upon deformation
\citep{marsden1983}.

At each point of the reference membrane there are two orthogonal directions
called principle extension directions for which the deformation is maximal and
minimal. The corresponding deformed line elements along these directions
remain orthogonal and have a stretched length given by the so called extension
ratios $\exten_i$ measured in units of the undeformed line elements.

An infinitesimal material patch on the reference shape with area
$d\reffl=\sqrt{\refmetric}d\lt d\lp$ will deform into the area element
$d\fl=\sqrt{\metric}d\lt d\lp$ on the current shape. The surface dilation
$\surdil$ is therefore given by the product of the extension ratios
\begin{equation}
   \surdil \equiv \frac{d\fl}{d\reffl} = \sqrt{\frac{\metric}{\refmetric}} =
   \exten_1\exten_2 \,.
   \label{eq:surdil}
\end{equation}
All scalar deformation quantities can be expressed by the extension ratios or
equivalently by the eigenvalues of the Lagrangian strain tensor $\lagr_i$,
which measure the difference of the extension ratios from unity
\begin{equation}
   \lagr_i\equiv \frac{1}{2}\left(\exten_i{}^2-1\right) \,.
   \label{eq:scalar}
\end{equation}

\clearpage


\clearpage


\end{document}